\documentclass{aastex}
\usepackage{emulateapj5}
\usepackage{onecolfloatx}
\begin{document}

\twocolumn[

\def\tm{$\m@th^{^{\rm T\kern -.04em M}}$}

\def\et{{\it et al.}}
\def\cbr{\mbox{\scriptsize CMB}}
\def\gap{\stackrel{>}{_\sim}}
\def\lap{\stackrel{<}{_\sim}}
\def\TBD{\bf TBD}

\title{The Ku-band Polarization Identifier}


\author{J. O. Gundersen\altaffilmark{1} for the KUPID Collaboration\altaffilmark{2}}
\affil{Department of Physics\\ University of Miami, Coral Gables, FL 33146}


\begin{abstract}
The Ku-band Polarization Identifier (KUPID) will integrate a very low noise 12-18 GHz, correlation polarimeter onto the Crawford Hill seven meter, millimeter-wave antenna.  The primary components of the polarimeter will be built at the University of Miami and other key components, including the microwave horn and data acquisition system will be built at the University of Chicago and Princeton University.  This project will measure the Q and U Stokes parameters in regions near the north celestial pole, in regions of low galactic contamination, and in regions near the galactic plane.  The KUPID survey experiment makes use of many of the techniques employed in the Princeton IQU Experiment (PIQUE) that was developed by the members of this collaboration to detect CMB polarization at shorter wavelengths.  The KUPID experiment will be constructed in parallel and on the same timescale as the CAPMAP experiment (see Barkats, this volume) which is the follow-on experiment to PIQUE.  KUPID will observe on the Crawford Hill antenna from late spring until early autumn, while CAPMAP will observe during the lower water vapor months of late autumn until early spring.  
\end{abstract}
]
\altaffiltext{1} {gunder@physics.miami.edu}
\altaffiltext{2}{The KUPID collaboration consists of members from the University of Miami (E. Stefanescu), Princeton University (D. Barkats, S.T. Staggs, J. McMahon, P. Farese), the University of Chicago (M. M. Hedman, B. Winstein, D. Samtleben, Keith Vanderlinde, Collin Bischoff), and JPL (T. Gaier).
}


\section{Introduction}

In the last few years the cosmology community has come to realize that a wealth of information can be gleaned from measurements of the angular power spectra of cosmic microwave background (CMB) polarization as well as its correlation with the temperature power spectrum.  Detailed studies of CMB polarization may detect the gravitational waves from the inflationary epoch (e.g. Kamionkowski and Jaffe 2000), help isolate the peculiar velocity at the surface of last scattering (Zaldarriaga and Harari 1995), determine the nature of primordial perturbations (e.g. Spergel and Zaldarriaga 1997), and probe primordial magnetic fields (e.g. Kosowsky and Loeb 1996) as well as cosmological parity violation (e.g. Lue, Wang, and Kamionkowski 1999).  A number of other possibilities are reviewed in Kamionkowski and Kosowsky 1999. This realization has manifested itself in the large number of new experiments devoted to CMB polarization studies (e.g. Timbie and Gundersen, 2002).  At least 12 of these experiments represent low frequency ($<100$ GHz) polarization measurements that may ultimately be limited by foreground contamination from polarized galactic synchrotron (Kogut and Hinshaw 2000, Davies and Wilkinson 1999), point sources, or perhaps the anomalous emission - hypothesized to be  "spinning dust" (Draine and Lazarian 1998a, 1998b). The level to which these foreground contaminants will limit the CMB polarization measurements is not well known (Tegmark et al. 1999).  The need for a low frequency survey to characterize the polarized emissions has been emphasized on many occasions (e.g. Cecchini et al. 2002).   The research objectives of the Ku-band Polarization Identifier (KUPID) directly address this need and go beyond.

\section{Observation Objectives}

KUPID will be configured as a polarimeter to measure both Q and U Stokes parameters, and it will also measure the continuum brightness temperature.  When the latter measurements are spatially differenced they can also be used as a differential radiometer to measure $\Delta T$.  With these capabilities, this instrument will be able to perform a wide variety of studies that span the interstellar medium to cosmology.  The primary research objectives include:\\
$\bullet$ Survey the polarized component of galactic synchrotron radiation\\
$\bullet$ Characterize the anomalous foreground emission - hypothesized to be "spinning dust"\\
$\bullet$ Measure CMB polarization, if foregrounds are not too limiting\\
$\bullet$ Perform follow-up measurements of interesting regions identified by the WMAP satellite\\

\subsection{Synchrotron} 

The bane of low frequency CMB polarization studies will likely be polarized Galactic synchrotron emission; however, no one knows the level of this contamination at the frequencies and angular scales of interest to CMB studies.  A summary of some of the existing observations derived from both polarization and temperature observations is presented in Figure 1.  When possible, this figure uses data from near the North Celestial Pole (NCP) at the angular scales of interest.  All the upper limits derived from both temperature and polarization measurements are shown as upside down triangles.  Actual detections of polarization derived from low frequency surveys are shown as other symbols.  The line that represents synchrotron radiation is normalized to the value of 0.5 $\mu K^2$ obtained by the WMAP satellite (Kogut et al. 2003) at 41 GHz.  This value is extrapolated to other frequencies using $T\propto\nu^{-2.9}$.  The other WMAP points in Figure 1 are obtained from the rms of the temperature anisotropy data in a $3^{\circ}$ radius cap centered on the NCP as derived from the maximum entropy method synchrotron maps (Bennett et al. 2003a).  A K-band (22 GHz) version of one of these NCP synchrotron maps is shown in Figure 3.  These points represent the maximum expected polarized synchrotron signal at these frequencies.  There are several uncertainties in extrapolating to other frequencies.  First, Faraday rotation can depolarize the low frequency measurements.  Second, there is uncertainty in the spectral index of the synchrotron.  The spectral index varies as a function of position on the sky (Reich \& Reich 1988) and is believed to steepen at higher frequencies due to a knee in the cosmic ray spectrum (Rybicki \& Lightman 1979).  The combination of these uncertainties makes the extrapolation to other frequencies quite suspect.  A Ku-band polarization survey will measure the normalization and slope of the synchrotron brightness spectrum and thus minimize many of these uncertainties.  In particular it will measure the polarized synchrotron in the region of interest for CAPMAP.

\begin{figure}[t]
\epsscale{1.0}
\plotone{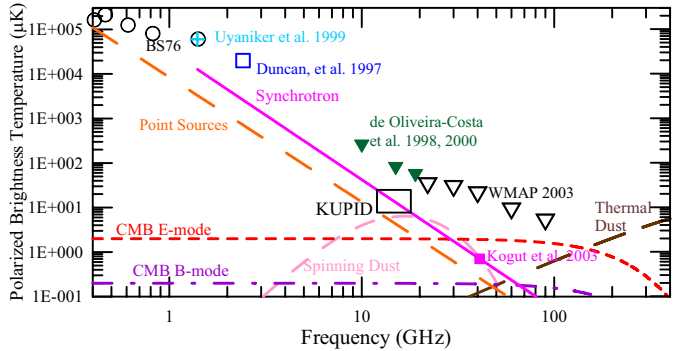}
\caption{\small Estimated brightness spectra of foreground contaminants and CMB signal as well as measured upper limits (upside-down triangles) and polarization detections (other symbols).  These foreground and CMB signal estimates are relevant to angular scales of $0.2^{\circ}-0.5^{\circ}$ and regions typical of the NCP.   The various foreground contaminants change as a function of galactic latitude (e.g. de Oliveira-Costa et al. 2000) and are expected to vary as a function of angular scale (Tegmark et al. 2000).  KUPID will operate at 12-18 GHz and will primarily map polarized synchrotron as shown by the box.  Details regarding the estimated slopes and normalizations of the various spectra are provided in the text.}
\label{fig:polspec}
\end{figure}

In addition to the slope and normalization of the synchrotron brightness spectrum, the slope and normalization of the polarized synchrotron angular power spectrum is of particular interest to CMB studies.  For many of the same reasons mentioned above, there is little known about angular power spectrum of polarized synchrotron at higher frequencies and at the high galactic latitudes of interest to CMB studies.  Recent studies from Baccigalupi et al. 2001 estimate the slope and normalization of either the E or B-mode angular power spectra with:
 \begin{equation}
C_{\ell}^{E,B}=(1.2\pm0.8)\times 10^{-9}(\ell /450)^{-1.8\pm0.3}(\nu_{GHz}/2.4)^{-5.8} K^2.
\end{equation}
This result is based on low and mid-latitude surveys (Duncan et al. 1997, 1999 and Uyaniker et al. 1999) at 1.4, 2.4 and 2.7 GHz as well as the results of Brouw and Spoelstra 1976 (BS76) on large angular scales.  The resulting power spectrum is similar to the results of Tucci et al. 2000.  The mean value of the normalization and slope are used for the synchrotron angular power spectrum shown in Figure 2 for an observation frequency of 15 GHz.  Based on these results the synchrotron spectrum (whether E or B) is expected to dominate over the CMB polarization power spectra and other foreground angular power spectra in KUPID's multipole band.  In addition to the quoted uncertainties in this determination of the normalization and slope of the synchrotron angular power spectrum, there are uncertainties in the extrapolation to higher frequencies, as mentioned in the discussion above.  Finally there are several caveats that need to be made about the validity of the Baccigalupi et al. 2001 results at high galactic latitudes, and there are also concerns regarding source subtraction (of HII regions) for the surveys used in the galactic plane.  KUPID's high signal/noise measurements at high and low galactic latitudes will reconcile many of these uncertainties.

\begin{figure}[t]
\epsscale{1.0}
\plotone{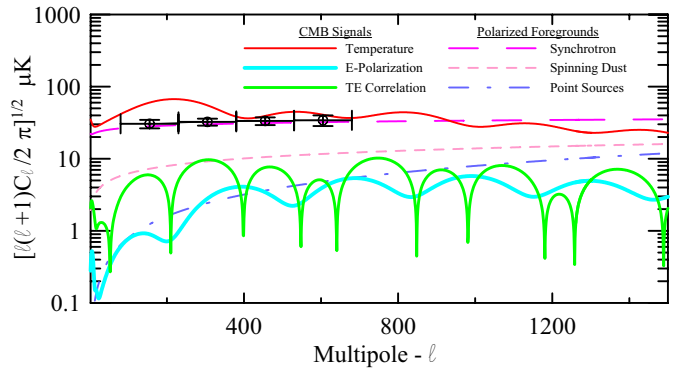}
\caption{\small Angular power spectra of the CMB signals compared to the angular power spectra of galactic and extragalactic signals at 15 GHz.  The CMB angular power spectra are calculated using CMBFAST (Seljak and Zaldarriaga 1996) assuming parameters derived from the WMAP data (Bennett et al. 2003b).  The slope and normalization of the spinning dust is derived from Tegmark et al. 2000.  The slope and normalization of the synchrotron and point source spectra are derived from Baccigalupi et al. 2001.  The estimated error bars on the synchrotron spectrum are based on 170 hours of observations in a $\delta>87^{\circ}$ cap.}
\label{fig:pow15}
\end{figure}

While polarized synchrotron radiation may be the bane of CMB polarization studies, it is a useful tool to study galactic magnetism (Beck et al. 1996).  Polarized synchrotron radiation is one of the key observational tracers of the plane-of-sky component of the galactic magnetic field $B_{\perp}$ (as opposed to the line-of-sight component $B_{\parallel}$).  The direction of polarization is perpendicular to $B_{\perp}$ while the strength of the 
polarization is proportional to $B^2_{\perp, u}/B^2_{\perp, t}$, where$B_t^2=B_u^2+B_r^2$ and $B_t, B_u, B_r$ are the total, uniform, and random components of the magnetic field (Ginzburg \& Syrovatskii 1965).  There is some controversy between estimates derived from radio frequency measurements of polarized galactic synchrotron radiation 
 (Spoelstra, 1984), which suggest that $B_u$ is slightly less than half the energy in fluctuations, and those derived from pulsar rotation measure data (Rand \& Kulkarni 1989 and Ohno \& Shibata 1993) which suggest that less than 10$\%$ of the energy density is carried by the uniform component.  Since KUPID will make very sensitive measurements of the polarized synchrotron radiation at frequencies where Faraday effects are small, it will provide estimates of the mean and random components of the galactic magnetic field.  This type of data will be useful in constraining various models that attempt to explain the origin and maintenance of galactic magnetic fields (Zweibel \& Heiles 1997).

\begin{figure}[t]
\epsscale{1.0}
\plotone{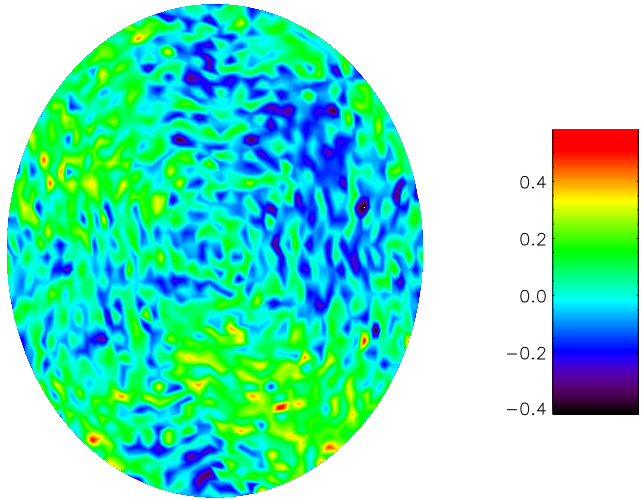}
\caption{\small A $\delta>87^{\circ}$ map of the synchrotron component derived from WMAP K-band data.  Units are in mK. (Bennett et al. 2003b)}
\label{fig:kband}
\end{figure}

\subsection{Anomalous Emission - (Spinning Dust?)}

 The second observation objective involves the characterization of the anomalous foreground emission that has been detected by five independent measurements including COBE (Kogut et al. 1996a, 1996b), Saskatoon (de Oliveira-Costa et al. 1997), OVRO (Leitch et al. 1997), the 19 GHz survey (de Oliveira-Costa et al. 1998, Boughn et al. 1992) and Tenerife (Gutierrez et al. 2000, de Oliveira-Costa et al. 1999 and Mukherjee et al. 2000).  This foreground is spatially correlated with 100 $\mu m$ dust emission but has a spectrum that rises towards lower frequencies.  This rise is incompatible with thermal dust emission and was initially attributed to free-free emission (Kogut et al. 1996a, 1996b).  Draine and Lazarian (1998a, 1998b) argued against the free-free hypothesis based on energetics and forwarded a hypothesis based on spinning (as opposed to vibrational) dust mission.  A cross-correlation analysis of the low frequency surveys with the Wisconsin H-Alpha Mapper 
(WHAM) data show little correlation, thus supporting the spinning dust hypothesis over the free-free component that is traced by the WHAM data (de Oliveira-Costa et al. 2000).  The brightness spectrum of the spinning dust shown in Figure 1 is normalized to the rms of the 100 $\mu$m dust emission (Wheelock et al. 1994) in the NCP region scaled to 15 GHz using 50 mK(MJy/sr)$^{-1}$ (de Oliveira-Costa et al. 2000) and assumes 5$\%$ polarization (Lazarian and Draine 2000).  Since the hypothesized spinning dust spectrum is predicted to peak in the 15-20 GHz regime, this experiment is ideal for measuring the polarized component (or lack thereof) of the anomalous emission.  As with the brightness spectrum, little is known about the angular power spectrum of the anomalous emission.  An estimate of this anomalous emission's angular power spectrum (Tegmark et al. 2000) is shown in Figure 2.  While the synchrotron emission is expected to dominate over the anomalous emission, the fact that they have different brightness spectra and the fact that the anomalous emission correlates with far-IR measurements will simplify the separation of these two sources.

\subsection{CMB Polarization}

 If neither the anomalous emission nor the synchrotron radiation is a dominant polarized foreground, then this experiment is well suited for studies of CMB polarization since it will be designed to have a high sensitivity and low systematic effects as discussed below.  Precision measurements of the CMB polarization angular power spectra will not only be the best independent confirmation of temperature anisotropy measurements, but they will also encode complementary information (e.g., Kosowsky 1999).  They will break degeneracies inherent in the parameter estimation derived from the temperature angular power spectrum, and they will constrain parameters that are not well constrained by temperature anisotropy measurements alone (e.g., Hu and White 1997).  While the signal-to-noise ratio of the temperature anisotropy measurements is typically a factor of 10-20 higher, the polarization signal is considered to be a ``cleaner" signal.  Phenomena such as the integrated Sachs-Wolfe effect and the Sunyaev-Zeldovich (SZ) effect, that occur between the current epoch (z=0) and the last scattering surface (at z=1000), affect the temperature anisotropy but not the polarization.  This fact can be used to disentangle low redshift ($z<1000$) effects from primordial phenomena.  In addition the polarization spectra are more sensitive to some cosmological parameters (such as the optical depth $\tau$) than the temperature spectrum.  Thus, even a lower signal to noise measurement on polarization can yield a better constraint on the optical depth.  Finally, the combined precision measurements of the temperature spectrum, the E and B mode polarization spectra and the T-E correlation spectrum will aid in the separation of the relative contributions from scalar, vector and tensor fluctuations (Hu and White 1997) - a goal that is unattainable with temperature measurements alone.  The quantification of the tensor contribution would be an amazing achievement since it would represent a measure of the stochastic background of primordial gravity waves.  A gravity wave background is the most direct and irrefutable evidence for an inflationary epoch at the very earliest stage of the Universe (Turner 1997).  While this experiment will not likely have the sensitivity to probe the B-mode polarization alone, it will determine the limiting level of foreground contaminants at low frequencies.  This information will be very important in the design of future experiments aimed at measuring B-mode polarization. 

\subsection{WMAP Follow-up}

The Wilkinson Microwave Anisotropy Probe (WMAP) satellite has made full sky temperature maps at five frequencies between 22 and 90 GHz with angular resolution that varies from $0.93^{\circ}- 0.23^{\circ}$, respectively (see http://map.gsfc.nasa.gov).  These maps encode a great deal of information about the CMB temperature anisotropy, the temperature-polarization correlation, and the galactic foregrounds.  This treasure trove of information has helped to answer many questions and has led to more refined questions regarding phenomena associated with both cosmology and the interstellar medium.  Many of these questions will benefit from the enhanced frequency coverage, angular resolution, and polarization sensitivity of experiments such as KUPID.  For instance, the WMAP satellite has identified the regions with the lowest synchrotron brightness temperature.  These regions will be prime targets for follow-up CMB polarization studies since they will also likely have the lowest polarized foreground emission.  Part of KUPID's observing time will be devoted to studies of these low foreground regions.

\section{Instrument Design}

One of the primary challenges of CMB polarization studies is the minimization of systematic effects.  This is a difficult task since the predicted polarization signal is a factor of 20-60 below the noise level that a single detector can obtain in one second and a factor of 100 million below ambient foreground temperatures.  The polarimeter, the telescope and the observation strategy are all designed to minimize the various systematic effects, some of which include polarized ground emission, correlated crosstalk, and polarization-dependent emission/absorption in the optics.  Both PIQUE (Hedman et al. 2001, 2002) and CAPMAP (Barkats, this volume) have demonstrated the ability to reach noise levels that are within a factor of 2-3 of the expected CMB polarization signal - without being limited by these types of systematic effects.  The KUPID experiment that will take advantage of this experimental heritage.

\begin{figure}[t]
\epsscale{1.0}
\plotone{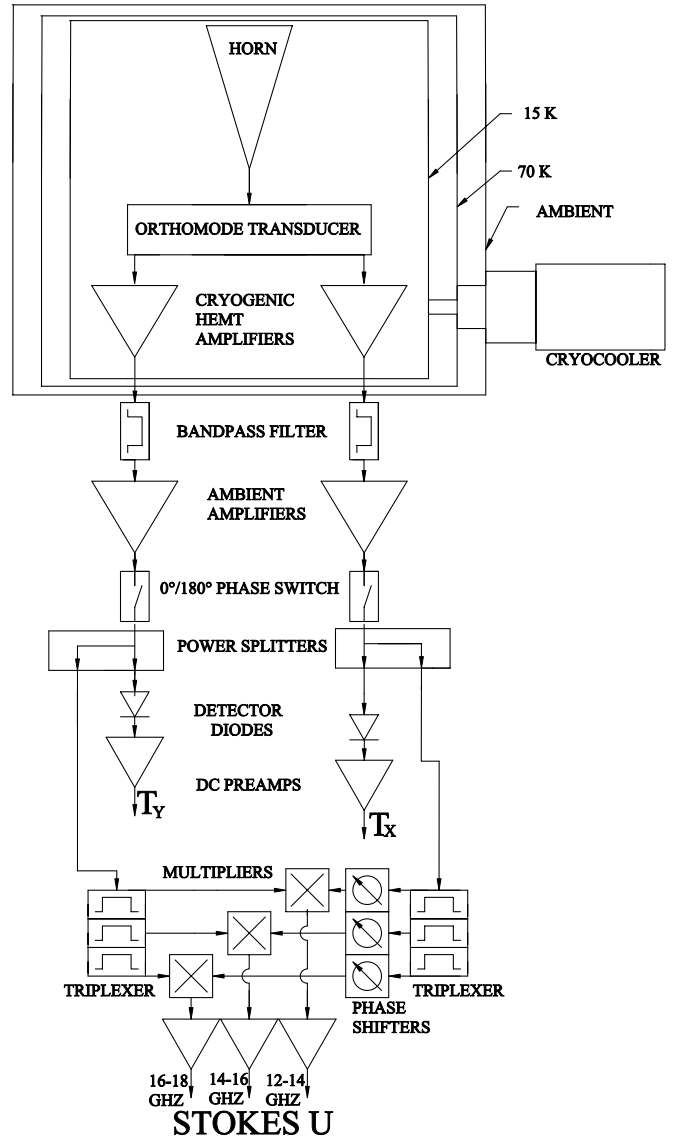}
\caption{\small Operational schematic of the Ku-band correlation polarimeter}
\label{fig:kuschem}
\end{figure}

\subsection{Polarimeter Design}

The primary instrument that requires development for KUPID is a phase switched correlation polarimeter (e.g. Duncan et al. 1995) that operates in the 12-18 GHz (Ku) band.  A diagram of the KUPID polarimeter is shown in Figure 4.  The main advantage of a correlation polarimeter is that it directly measures the Q and U Stokes parameters rather than taking the difference between two large signals like the MAP experiment and many bolometric-based experiments will do.  In so doing, gain fluctuations in the front-end amplifiers only multiply the small polarization signal rather than the system temperature as in the total power case.  A phase switch is introduced into one arm of the polarimeter, and this reverses the sign of the multiplier output.  The phase switching and synchronous demodulation obviates the small offset term that is inherent in the multiplier output, and the switching is performed at a frequency (several kHz) that is well above the 1/f knee of the cryogenic amplifiers.  Other similarities with the PIQUE and CAPMAP systems includes the use of low noise cryogenic amplifiers, an ambient temperature amplifier and a power divider that allows for the detection of the total intensity in each arm - in this case the intensity of the left and right circular polarizations.   The triplexers in the backend of KUPID split the 12-18 GHz into 3 equal sub-bands.  This allows for frequency discrimination among various astrophysical sources, and it also enables easier phase matching since the fractional bandwidth is smaller.

There are several differences between the KUPID polarimeter and the PIQUE/CAPMAP  polarimeters.  First, KUPID is a homodyne receiver rather than a heterodyne receiver.  This significantly simplifies the construction and testing of the polarimeter, and it reduces the overall cost.  KUPID will not need a local oscillator or an associated magic tee and phase shifter, and perhaps most importantly, there'll be no cryogenic mixer.  The primary source of reduced effective bandwidth of PIQUE's Q-band and W-band polarimeters has been the lack of flatness in the cryogenic mixers' conversion loss.  The desired goal for KUPID will realize a large fraction ($>90\%$) of its potential 6 GHz bandwidth.  The second difference is that the phase switching is performed in-line rather than in the local oscillator signal as is done for CAPMAP.  This technique has worked successfully for the MAP satellite.  The third difference between the polarimeters is that KUPID's lens is quite large (approximately 50 cm in diameter) and it will initially be at ambient temperature.  

The Ku-band is chosen for a number of science reasons: 1) As shown in Figure 1, polarized synchrotron emission is expected to be sizable at these frequencies, 2) very little mapping of the diffuse polarization has been performed at these frequencies, 3) this band is complementary to the higher frequency bands of the MAP satellite that extend from 22 to 90 GHz, and 4) this band straddles the peak of the hypothesized spinning dust foreground.  There are also several technical/practical reasons to chose Ku-band: 1) A simple homodyne system can be incorporated with no waveguide components (except the horn, the OMT and the quadrature hybrid), 2) similar components used in the intermediate frequency (IF) portions of PIQUE/CAPMAP can be also incorporated into this system, and 3) very low noise, cryogenic amplifiers are available in the Ku-band from the National Radio Astronomy Observatory.

\subsection{Sensitivity}

The sensitivity of a correlation receiver is given by $\sqrt{2}T_{sys}/\sqrt{\Delta\nu}$  (Kraus 1986), where $T_{sys}$ is the system temperature and $\Delta\nu$ is the bandwidth and this is often expressed in terms of $\mu K\sqrt{sec}$.  The various factors that contribute to the system temperature include the CMB itself (2.7 K), the atmosphere (3-6 K at 12-18 GHz), the loss in the lens, horn and waveguide (estimated to be $<0.1$ dB), and the receiver noise temperature (7 K) which is dominated by the cryogenic amplifiers.  These factors give a system temperature of $\sim 20$ K, and when this is combined with a 6 GHz bandwidth it results in a sensitivity of 370 $\mu$K$\sqrt{sec}$ in the measurement of either the Q or U Stokes parameter.  

\subsection{Antenna}

KUPID will use the 7-meter millimeter wave antenna at Crawford Hill (Chu et al. 1978), and it will share observing time with CAPMAP.  There are a number of reasons that this antenna is preferred over other possible antennas:

1) First and foremost, the KUPID experiment and the CAPMAP experiment share many of the same personnel and hardware; thus it is prudent, from a budgetary and personnel perspective, to site both experiments at the same antenna.  In this plan, CAPMAP would observe from late autumn until early spring when the precipitable water vapor is lowest, and KUPID would observe during the other months.

2) The Crawford Hill telescope is an offset Cassegrain antenna that resides on an alt-az mount. Its characteristics are well tailored for the measurement of polarization.  The off-axis structure and the low surface error on the mirror ($<$ 100 $\mu$m) afford low sidelobe and cross-polarization levels.  No other telescope of similar size (CSO, HHT, SEST, FCRAO, Owens Valley 10-m, JCMT) offers the combination of these features.  In addition very few, if any, of these other antennas would allow the extended observing time that is required.

3) The FWHM beam size will be 0.2$^{\circ}$ at 15 GHz.  This is well matched to PIQUE's 90 GHz measurements (0.24$^{\circ}$ FWHM), and it is also matched to WMAP's highest frequency (90 GHz) channels which have a 0.2$^{\circ}$ FWHM.  A much larger off-axis telescope (such as the GBT in the extreme case) would allow increased angular resolution; however, we would not be able to map large regions (10's of square degrees) in a 
reasonable time at the requisite sensitivity level.  A small, off-axis telescope (such as that used for PIQUE) does not provide the angular resolution necessary for straightforward comparisons with WMAP and other higher angular resolution CMB temperature and polarization experiments.

\subsection{Observing Strategy}

The initial survey will focus on the NCP region ($\delta>87^{\circ}$) because COMPASS, PIQUE and now CAPMAP have focused their higher frequency polarization observations near the NCP.  The NCP region is chosen because it affords long continuous observations of a very small patch of sky without the need to change elevation.  The long, deep observations are required to observe the very small polarization signal.  A region corresponding to declinations larger than 87$^{\circ}$ can be measured by KUPID to a precision of 10 $\mu$K/beam in 170 hours.  An estimate of the uncertainties in the measured synchrotron angular power spectrum are shown in Figure 2.  This measurement will more than encompass the regions that have already been observed by any of the above experiments.  If no polarization is observed at this precision, then it would support the conclusion that neither spinning dust nor synchrotron is interfering with the higher frequency measurements since the brightness temperature spectrum drops precipitously at higher frequencies.  In this case KUPID would continue observing in the $\delta>87^{\circ}$ region and push to 5 $\mu$K/beam precision and this would take an additional 510 hours of observing.  At this level the CMB polarization signal is expected to become significant.  In the event that significant polarization is observed at the 10 $\mu$K/beam level, then KUPID would perform a larger survey near the NCP.  A survey region of $\delta>80^{\circ}$ would take approximately 420 hours or 17.5 days.  This low frequency survey will be invaluable to future CMB polarization surveys of the region.  Both the deep survey and the extended survey can be completed in the first three month season of observations.  This takes into account the fact that previous polarization experiments (PIQUE, COMPASS and POLAR) only observed for about 25$\%$ of the time due to poor weather and occasional experimental downtime.

The observing strategy of the second observing season will be dictated by what is observed in the first season.  If no foreground contamination is observed, then these seasons will focus on deep CMB polarization surveys.  If significant foreground contamination is observed, then KUPID will perform follow-up observations in interesting regions observed by other polarization experiments such as MAXIPOL.  Another goal would be to establish the foreground contamination level in regions with low foreground levels, such as the North Galactic Pole.  In addition, a galactic plane survey covering regions from galactic longitudes 0$^{\circ}$ to 250$^{\circ}$ and galactic latitudes of -5$^{\circ}$ to 5$^{\circ}$ could be conducted to a precision of 100 $\mu$K/beam in approximately 151 hours.  This precision should be adequate since existing surveys (e.g. Uyaniker et al. 1998) have peak-to-peak variations in Q and U of 500 $\mu$K at 1.4 GHz.  If this is scaled to 14 GHz using $T\propto\nu^{-3}$, the peak-to-peak values near the plane of the galaxy will be on order of 500 $\mu$K.  

\subsection{Calibration}

The calibration of the polarimeter itself will be established in the lab using a second OMT, in place of the horn, with variable temperature cold loads at its two inputs.  The two OMTs are rotated 45$^{\circ}$ with respect to each other such that when the two loads are at different temperatures, it generates a nonzero Q (or U) at the output of the correlators.  In the process of varying the two load temperatures, the gain of the polarimeter can be established. The polarized flux density scale of the telescope can be calibrated using a bright, polarized astronomical source such as the Crab Nebula.  The Crab has been well characterized at these frequencies (e.g. Mayer and Hollinger 1968).  The antenna temperature of the total linear polarization $P=\sqrt{Q^2+U^2}$ of the Crab is expected to be about 100 mK at 18 GHz for KUPID's $0.2^{\circ}$ FWHM beam.  A signal/noise of well over 100 can be achieved in one second of integration.  Measurements of the Crab's Stokes parameters  can be completed over a wide range of parallactic angles ($PA_{AZ}$) yielding signals:
\begin{equation}
Q_{out}=P_{crab}\cos(2(PA_{src}-PA_{AZ}))
\end{equation}
\begin{equation}
U_{out}=P_{crab}\sin(2(PA_{src}-PA_{AZ}))
\end{equation}
for an idealized polarimeter/feed system, where $PA_{src}=0.5\tan^{-1}(U_{src}/Q_{src})$ is the position angle of the source.  Deviations from this ideal behavior can be attributed to non-ideal components in the polarimeter/feed system which can modify the amplitude and phase of the incident Stokes parameters.  This non-ideal behavior causes coupling between the measured Stokes parameters (I, Q, U) and this coupling can be characterized in terms of a Mueller matrix (Tinbergen 1996) for each component in the polarimeter chain.  A high signal/noise measurement of the Crab or another bright calibration target will enable the precision determination of the various terms in the Mueller matrix, and these can in turn be used to correct for the instrumental effects.  A thorough discussion of this technique is given in Heiles et al. 2001 and Heiles 2001.  

\subsection{Timeline}

Year 1 (8/02-7/03) - During the first year the primary goals involve the design, construction, and testing of the polarimeter.  The telescope-polarimeter interface will be established, and an ambient version of the complete polarimeter will be tested.  A replica of the CAPMAP data acquisition card will be constructed and tested.  Towards the end of the year the cryostat will be completed and the cryocooler and compressor will be delivered.  The month of June will be used to perform cryogenic tests of the polarimeter and once it is operating well, it will be moved to the Crawford Hill telescope.

	Year 2 (8/03-7/04) - The months of August, September and October will be spent observing a small cap near the NCP.  KUPID will fit into one of the four cryostat slots alloted to CAPMAP until all four CAPMAP cryostats are completed and deployed (12/03).  These months are chosen to complement CAPMAP's observing season which is October through May.  The rest of the year will be dominated by data analysis from the first observing season.  The other months will also be used to incorporate modifications to the polarimeter system.  

	Year 3 (8/04-7/05) - The months of August-October will be used to perform the observations, and the rest of the year will be used to conclude the data analysis.  Based on the results from Year 2, we will determine the observation strategy for Year 3 as discussed in the Observation Strategy section.

\acknowledgments
This work was supported in part by NSF grant AST-0206241.  JG is grateful to V. Shenouda for her assistance in extracting Figure 3 from the WMAP archive.


\clearpage







\clearpage


\begin{thebibliography}{}

\bibitem[Baccigalupi et al. 2002]{bacc02}Baccigalupi, C., De Zotti, G., Burigana, C., and Perrotta, F.  2002,  Astrophysical Polarized Backgrounds, AIP Conference Proceedings, Cecchini, S., Cortiglioni, S., Sault, R., and Sbarra, C. eds., 609, 84

\bibitem[Beck et al. 1996]{beck96}Beck, R., Brandenburg, A., Moss, D., Shukurov, A., and Sokoloff, D.  1996,  Annu. Rev. Astron. and Astrophys., 34, 155

\bibitem[Bennett et al. 2003a]{ben03a}Bennett, C. L., Hill, R. S., Hinshaw, G., Nolta, M. R., Odegard, N., Page, L., Spergel, D. N., Weiland, J. L., Wright, E. L., Halpern, M., Jarosik, N., Kogut, A., N., Limon, M., Meyer, S. S., Tucker, G. S., and Wollack, E.  2003a, submitted to ApJ

\bibitem[Bennett et al. 2003b]{ben03b}Bennett, C. L., Halpern, M., Hinshaw, Jarosik, N., Kogut, A., Limon, M., Meyer, S. S., Page, L., Spergel, D. N., Tucker, G. S., Wollack, E., Wright, E. L., Barnes, C., Greason, M. R., Hill, R. S., Komatsu, E., Nolta, M. R., Odegard, N., Peiris, H. V., Verde, L., and Weiland, J. L.  2003b, submitted to ApJ


\bibitem[Boughn et al. 1992]{bou92}Boughn, S.P., Cheng, E.S., Cottingham, D.A., and Fixsen, D.J.  1992,  ApJ, 391, L49

\bibitem[Brouw \& Spoelstra 1976]{brouw76}Brouw, W.N., \&
Spoelstra, T.A.T. 1976, A\&AS, 26, 129

\bibitem[Cecchini et al. 2002]{cecc02}Cecchini, S., Cortiglioni, S., Sault, R., and Sbarra, C. eds.  2002, Astrophysical Polarized Foregrounds, AIP Conference Proceedings, Vol. 609.

\bibitem[Chu et al. 1978]{chu78}Chu, T. S., Wilson, R. W., England, R. W., Gray, D. A., and Legg, W. E.  1978,  The Bell System Technical Journal, 57, 5, 1257.

\bibitem[Davies \& Wilkinson 1999]{davwil99}Davies, R.D., and Wilkinson, A.  1999,  Microwave Foregrounds, ASP Conference Series, A. de Oliveira-Costa and M. Tegmark, eds., 181, 77

\bibitem[de Oliveira-Costa et al. 1997]{oliv97}de Oliveira-Costa, A., Kogut, A., Devlin, M.J., Netterfield, C.B., Page, L.A., and Wollack, E.J.  1997,  ApJ, 482, L17.

\bibitem[de Oliveira-Costa et al. 1998]{oliv98}de Oliveira-Costa, A., Tegmark, Page, L.A., and Boughn, S.P.  1998, ApJ, 509, L9.

\bibitem[de Oliveira-Costa et al. 1999]{oliv99}de Oliveira-Costa, A., Tegmark, M., Gutierrez, C.M., Jones, A.W., Davies, R.D., Lasenby, A.N., Rebolo, and Watson, R.A.  1999, ApJ, 527, L9.

\bibitem[de Oliveira-Costa et al. 2000]{oliv00}de Oliveira-Costa, A., Tegmark, M., Finkbeiner, D.P., Davies, R.D., Gutierrez, C.M., Haffner, L.M., Jones, A.W., Lasenby, A.N., Rebolo, R., Reynolds, R.J., Tufte, S.L., and Watson, R.A.  2000, astro-ph/00110527.

\bibitem[Draine \& Lazarian 1998a]{drlaz98a}Draine, B.T., and Lazarian, A.  1998a, ApJ, 494, L19.

\bibitem[Draine \& Lazarian 1998b]{drlaz98b}Draine, B.T., and Lazarian, A.  1998b, ApJ, 508, 157.

\bibitem[Draine \& Lazarian 1999]{dra99}Draine, B. T., \& Lazarian, A.
1999, in Microwave Foregrounds, eds. A. de Oliveira-Costa
\& M. Tegmark (ASP: San Francisco), 133

\bibitem[Duncan et al. 1995]{dun95}Duncan, A.R., Stewart, R.T., Haynes, R.F., and Jones, K.L.  1995,  MNRAS, 277, 36.

\bibitem[Duncan et al. 1997]{dun97}Duncan, A.R., Haynes, R.F., Jones, K.L., and Stewart, R.T.  1997, MNRAS, 291, 279.

\bibitem[Duncan et al. 1999]{dun99}Duncan, A. R., Reich, P., Reich, W., and Furst, E.  1999, A\&A, 350, 447.

\bibitem[Ginzburg \& Syrovatskii 1965]{ginz65}Ginzburg, V.L. and Syrovatskii, S.I.  1965,  Annu. Rev. Astron. Astrophys., 3, 297.

\bibitem[Gutierrez et al. 2000]{gut00}Gutierrez, C.M., Rebolo, R., Watson, R.A., Davies, R.D., Jones, A.W., and Lasenby, A.N.  2000, ApJ, 529, 47.

\bibitem[Hedman et al. 2001]{hed01}Hedman, M. M., Barkats, D., Gundersen, J. O., Staggs, S. T., and Winstein, B. 2001, 548, L114.

\bibitem[Hedman et al. 2002]{hed02]}Hedman, M. M., Barkats, D., Gundersen, J. O., McMahon, J. J., Staggs, S. T., and Winstein, B. 2002, ApJ, 573, L73.


\bibitem[Heiles 2001a]{hei01a}Heiles, C.  2001, astro-ph/0107327.

\bibitem[Heiles et al. 2001]{hei01b}Heiles, C., Perillat, P., Nolan, M., Lorimer, D., Bhat, R., Ghosh, T., Lewis, M., O'Neil, K., Salter, C., Stanimirovic, S.  2001, astro-ph/0107352.

\bibitem[Hu \& White 1997]{huwhi97}Hu, W. \& White, M. 1997, NewA, 2, 323

\bibitem[Kamionkowski \& Kosowsky]{kamkow99}Kamionkowski, M., and Kosowsky, A.  1999, Ann. Rev. Nucl. Part. Sci., 49, 77


\bibitem[Kamionkowski et al. 1997]{kamkow97}Kamionkowski, M., 
Kosowsky, A., \& Stebbins, A. 1997, PRD, 55, 7368



\bibitem[Kamionowski \& Jaffe 2000]{kamjaf00}Kamionkowski, M., and Jaffe, A.H.  2000, "Detection of Gravitational Waves from Inflation," astro-ph/0011329

\bibitem[Kogut et al. 1996a]{kog96a}Kogut, A., Banday, A.J., Bennett, C.L., Gorski, K.M., Hinshaw, and Reach, W.T.  1996a, ApJ, 460, 1.

\bibitem[Kogut et al. 1996b]{kog96b}Kogut, A., Banday, A.J., Bennett, C.L., Gorski, K.M., Hinshaw, G., Smoot, G.F., and Wright, E.L.  1996b,  ApJ, 464, L5.

\bibitem[Kogut \& Hinshaw]{koghin00}Kogut, A. and Hinshaw, G.  2000, ApJ, 543, 530

\bibitem[Kogut et al. 2003]{kog03}Kogut, A., Spergel, D. N., Barnes, C., Bennett, C. L., Halpern, M., Hinshaw, G., Jarosik, N., Limon, M., Meyer, S. S., Page, L., Tucker, G. S., Wollack, E., and Wright, E. L.  2003, submitted to ApJ

\bibitem[Kosowsky \& Loeb]{kowloeb96}Kosowsky, A. and Loeb, A.  1996,  ApJ, 469, 1

\bibitem[Kosowsky 1999]{kos99}Kosowsky, A. 1999, NewA, 43, 147

\bibitem[Krauss 1986]{krauss}Krauss, J. D., 1986, Radio
Astronomy, (2d ed.; Powell, OH: Cygnus-Quasar Books)

\bibitem[Lazarian \& Draine]{laz00}Lazarian, A., and Draine, B.T.  2000, astro-ph/0003312.

\bibitem[Leitch et al. 1997]{leitch97}Leitch, E.M., Readhead, A.C.S., Pearson, T.J., and Myers, S.T.  1997,  ApJ, 486, L23.

\bibitem[Lue 1999]{lue99}Lue, A., Wang, L., and Kamionkowski, M.  1999,  Phys. Rev. Lett., 83, 1506

\bibitem[Mayer \& Hollinger]{may68}Mayer, C.H., and Hollinger, J.P.  1968, ApJ, 151, 53.

\bibitem[Mukherjee et al. 2000]{muk00}Mukherjee, P., Jones, A.W., Kneissl, R., and Lasenby, A.N.  2000,  astro-ph/0002305.

\bibitem[Ohno \& Shibata 1993]{ohno93}Ohno, H. and Shibata, S.  1993,  MNRAS, 262, 953.

\bibitem[Rand \& Kulkarni 1989]{rand89}Rand, R. J. and Kulkarni, S. R.  1989,  ApJ, 343, 760.

\bibitem[Reich \& Reich]{rr88}Reich, P. and Reich, W.  1988, A\&AS, 74, 7

\bibitem[Rybicki \& Lightman]{ryblight79}Rybicki, G.B. and Lightman, A.P, 1979, "Radiative Processes in Astrophysics," (Wiley and Sons, New York).

\bibitem[Seljak, U. \& Zaldarriaga]{sel96}Seljak, U. \& Zaldarriaga, M.  1996, ApJ, 469, 437

\bibitem[Spergel \& Zaldarriaga]{spergzal97}Spergel, D.N., and Zaldarriaga, M.  1997,  Phys. Rev. Lett., 79, 2180

\bibitem[Spoelstra 1984]{spoel84}Spoelstra, T.A.T.  1984,  A\&A, 135, 238.

\bibitem[Tegmark et al. 1999]{teg99}Tegmark, M., Eisenstein, D.J., Hu, W., and de Oliveira-Costa, A.  1999,  Microwave Foregrounds, ASP Conference Series, A. de Oliveira-Costa and M. Tegmark, eds., 181, 3.

\bibitem[Tegmark et al. 2000]{teg00}Tegmark, M., Eisenstein, D.J., Hu, W., and de Oliveira-Costa, A.  2000, ApJ, 530, 133.

\bibitem[Tegmark \& Zaldarriaga 2000]{tegzal00}Tegmark, M. \& 
Zaldarriaga, M.  2000, PRL, 85, 2240

\bibitem[Tinbergen 1996]{tin96}Tinbergen, J.  1996, "Astronomical Polarimetry," Cambridge University Press

\bibitem[Timbie \& Gundersen 2002]{tim02}Timbie, P.T., and Gundersen, J. O., AMiBA 2001: High-z Clusters, Missing Baryons, and CMB Polarization, L.W. Chen, C.P. Ma, K.W. Ng and U. L. Pen, eds., ASP CS-257, 235, 2002.

\bibitem[Tucci et al. 2001]{tuc01}Tucci, M., Carretti, E., Cecchini, S., Fabbri, R., Orsini, M., and Pierpaoli, E., 2001, submitted to Elsevier, astro-ph/0006387

\bibitem[Turner 1997]{turn97}Turner, M.S.  1997, Generation of Cosmological Large-scale Structure, ed. D.N. Schramm (Kluwer, Dordrecht)

\bibitem[Uyaniker et al. 1998]{uyan98}Uyaniker, B., Furst, E., Reich, W., Reich, P., and Wielebinski, R.  1998,  A\&S, 132, 401

\bibitem[Wheelock et al. 1994]{IRAS}
Wheelock, S. L., et al. 1994, IRAS Sky Survey Atlas: Explanatory
Supplement (Pasadena: JPL 94-11)

\bibitem[Zaldarriaga \& Harari 1995]{zalhar95}Zaldarriaga, M. \& Harari, D.D.  1995,  PRD, 52, 3276

\bibitem[Zweibel \& Heiles]{Zwei97}Zweibel, E.G. and Heiles, C.  1997,  Nature, 385, 131.

\end{thebibliography}
\end{document}